\documentclass[aps,prl,reprint,amsmath,amssymb,floatfix,superscriptaddress]{revtex4-2}
\usepackage{graphicx,bm,amsmath,amssymb,physics}
\usepackage[colorlinks=true,linkcolor=blue,citecolor=blue,urlcolor=blue]{hyperref}
\graphicspath{{Figures/}}
\newcommand{\UK}{U_{\mathrm K}}
\newcommand{\UI}{U_{\mathrm I}}
\newcommand{\EN}{E_{\mathcal N}}
\newcommand{\Psym}{P_{\mathrm{ring}}}
\begin{document}
\title{Extracting Operational Entanglement from a Noise-Protected Many-Body System}
\author{Bhavay Tyagi}
\email{btyagi@uh.edu}
\affiliation{Department of Physics, University of Houston, Houston, Texas 77204, USA}
\author{Eric R. Bittner}
\email{ebittner@central.uh.edu}
\affiliation{Department of Physics, University of Houston, Houston, Texas 77204, USA}
\author{Kevin E. Bassler}
\email{bassler@uh.edu}
\affiliation{Department of Physics, University of Houston, Houston, Texas 77204, USA}
\affiliation{Department of Mathematics, University of Houston, Houston, TX}
\affiliation{Texas Center for Superconductivity, University of Houston, Houston, TX}
\date{November 19, 2026}
\begin{abstract}
Can noisy dynamics generate entanglement that stays useful at long times? Under collective dephasing, a twelve-spin kicked Ising ring evolves from a uniform product state toward an entangled stationary state that is maximally mixed in the sector fixed by the ring’s rotations and reflections. Gates and measurements within each half herald a spin pair that teleports an unknown qubit with a mean fidelity of 90.5\%. Simulations show that stronger noise correlations extend the time over which the ring yields pairs that beat the classical teleportation limit.
\end{abstract}
\maketitle

\begin{figure}[!t]
\includegraphics[width=\linewidth]{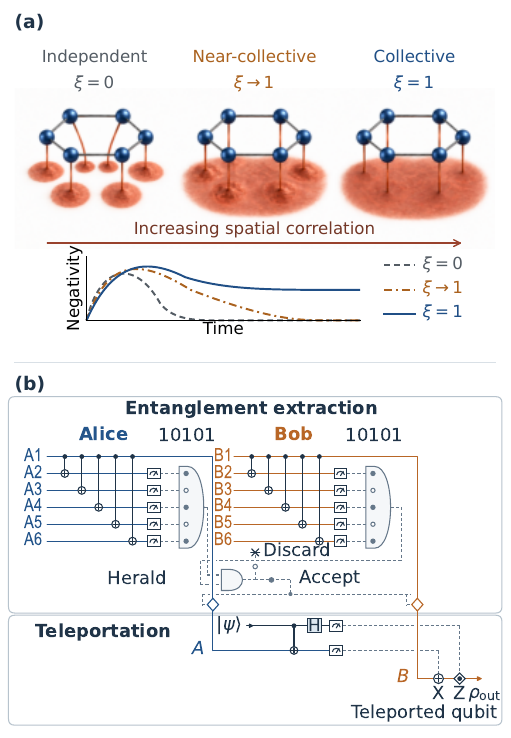}
\caption{\label{fig:model} (a) Independent and collective dephasing of the interacting ring (blue spins; red environment). Collective dephasing cannot distinguish spin configurations with the same total magnetization. Schematic negativity curves illustrate entanglement survival as $\xi\rightarrow1$. The ring-and-environment artwork in Fig.~\ref{fig:model}(a) was generated and edited using OpenAI's image generation tool. (b) Physical extraction and teleportation. Alice and Bob apply local CNOT gates and measurements within their halves, accepting the pair ($A$ and $B$) when both records are $10101$. Bell-state measurement and Pauli corrections teleport the unknown input $\ket{\psi}$ to $\rho_{\text{out}}$.}
\end{figure}

We traditionally treat noise as the enemy of coherence, and counter it either by isolating the system better or by correcting its errors~\cite{Acharya2023}. A complementary route asks what the environment can actually \emph{see}. Figure~\ref{fig:model} (a) illustrates this distinction. Independent noise channels resolve the coupled spin component at each site, whereas a collective field couples only to their sum. Spatial correlations in the environment therefore determine how well it can distinguish different spin configurations. Decoherence-free subspaces and noiseless subsystems exploit collective coupling to preserve quantum information~\cite{Zanardi1997,Lidar1998,Knill2000}. Correlated dephasing has also been measured in trapped-ion, superconducting and silicon-spin devices~\cite{Monz2011,vonLupke2020,Yoneda2023}. Our earlier work showed how correlated environments organize collective quantum dynamics~\cite{Tyagi2024,Bittner2025jcp,Bittner2026graph}. We now carry this question into an interacting many-body system. Can a correlated environment protect entanglement that the system generates intrinsically, starting from an unentangled state? What fixes its surviving amount, and can it be used to perform a quantum task?

Strong symmetries also organize dissipative dynamics into invariant sectors~\cite{BucaProsen2012}. For unital channels with non-Abelian on-site symmetries, maximally mixed invariant-sector states can be entangled and have distillable entanglement~\cite{Moharramipour2024}. We ask whether spatial symmetry can protect an entangled state that a driven ring prepares from a product state. To use that entanglement, we bring it onto two physical spins with gates and measurements within each half of the ring.

\emph{In this Letter} we show how a noisy driven Ising ring~\cite{Bertini2018} prepares an entangled state and supplies a physical pair for teleportation. The protecting symmetries rotate and reflect the ring, rather than act on each site as in Ref.~\cite{Moharramipour2024}. Starting from a uniform product state, interactions generate entanglement while collective dephasing keeps the state within the ring-symmetric sector. For twelve spins, the dynamics approaches the maximally mixed state within this sector. A circuit acting within each half then extracts a pair that teleports a qubit above the classical fidelity benchmark [Fig.~\ref{fig:model}(b)]. The protocol turns entanglement generated in a noisy many-body system into an accessible quantum resource.

\emph{Model and preparation.}---The one-period quantum channel $\mathcal M_\xi$ applies an Ising layer, a transverse kick, and accumulated dephasing, in that order.
\begin{align}
\UI&=\exp\!\left[-i\left(J\sum_{j=1}^{L}\sigma_j^z\sigma_{j+1}^z+h\sum_{j=1}^{L}\sigma_j^z\right)\right],\nonumber\\
\UK&=\exp\!\left[-ib\sum_{j=1}^{L}\sigma_j^x\right],\nonumber\\
\rho_{t+1}&=\mathcal M_\xi[\rho_t]\nonumber\\
&\equiv e^{\mathcal D_\xi}\!\left(\UK\UI\rho_t\UI^\dagger\UK^\dagger\right).
\label{eq:map}
\end{align}
Here $L$ is the number of spins (the system volume), and $\sigma_j^{x,y,z}$ are local Pauli operators acting on site $j$. The boundary is periodic, $\sigma^z_{L+1}=\sigma^z_1$, and one period is the unit of time. At self-dual points, \emph{space-time} duality permits exact calculations of selected spectral and entanglement observables of the coherent dynamics~\cite{Bertini2018,Bertini2019}. We use twelve spins with $J=\pi/4$, $b=-\pi/4$, and $h=1$, at a \emph{dual-unitary point}. The initial state is the uniform product $\rho_0=(\ket{+}\!\bra{+})^{\otimes L}$, where $\ket{+}=(\ket0+\ket1)/\sqrt2$. Appendix~\ref{App:A} proves convergence to the symmetry-selected stationary state under collective dephasing at these parameters.

We describe the accumulated dephasing by the Lindblad generator~\cite{Breuer2007},
\begin{equation}
\mathcal D_\xi[\rho]=\gamma\sum_{i,j=1}^{L}\Xi_{ij}\left(\sigma_i^z\rho\sigma_j^z-\tfrac12\{\sigma_j^z\sigma_i^z,\rho\}\right).
\label{eq:noise}
\end{equation}
Here $\sigma_i^z$ are the local dephasing operators and $\Xi_{ij}=(1-\xi)\delta_{ij}+\xi$ is the noise-correlation matrix. Its eigenvalues give the physical range $-1/(L-1)\le\xi\le1$ (Appendix~\ref{app:Log-neg}). Increasing $\xi$ strengthens the uniform collective noise mode while keeping the noise variance at each spin fixed. The noise is independent at $\xi=0$ and fully collective at $\xi=1$. Negative $\xi$ describes anticorrelated phase fluctuations on distinct sites. Averaging Gaussian white phase noise with $\gamma=0.1$ after the coherent layers gives $e^{\mathcal D_\xi}$ exactly.

To quantify the entanglement generated during evolution, we divide the ring into two contiguous equal halves and calculate the logarithmic negativity~\cite{Vidal2002},
\begin{equation}
\EN(t)=\ln\|\rho_t^{T_A}\|_1.
\label{eq:negativity}
\end{equation}
A positive value certifies entanglement between the two halves. The product input has $\EN(0)=0$.

\begin{figure}[!t]
\includegraphics[width=\linewidth]{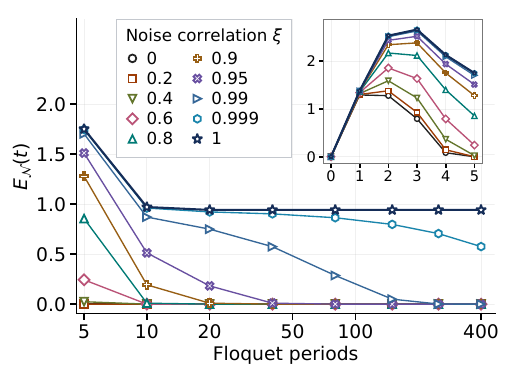}
\caption{\label{fig:negativity}Logarithmic negativity $\EN(t)$ across the equal bipartition of the twelve-spin ring for selected correlations from independent ($\xi=0$) to collective ($\xi=1$) dephasing. Markers indicate the periods at which $\EN(t)$ was evaluated. The inset shows periods $0$ to $5$ on a linear time axis. The main time axis is logarithmic.}
\end{figure}

\emph{Results.}---Interactions generate entanglement between the two halves from the product input. Figure~\ref{fig:negativity} shows a rapid buildup followed by correlation-dependent decay. Near the collective limit, entanglement survives longer as the noise becomes more correlated. At $\xi=1$, the trajectory approaches an entangled stationary state.

\begin{figure*}[t]
\includegraphics[width=\textwidth]{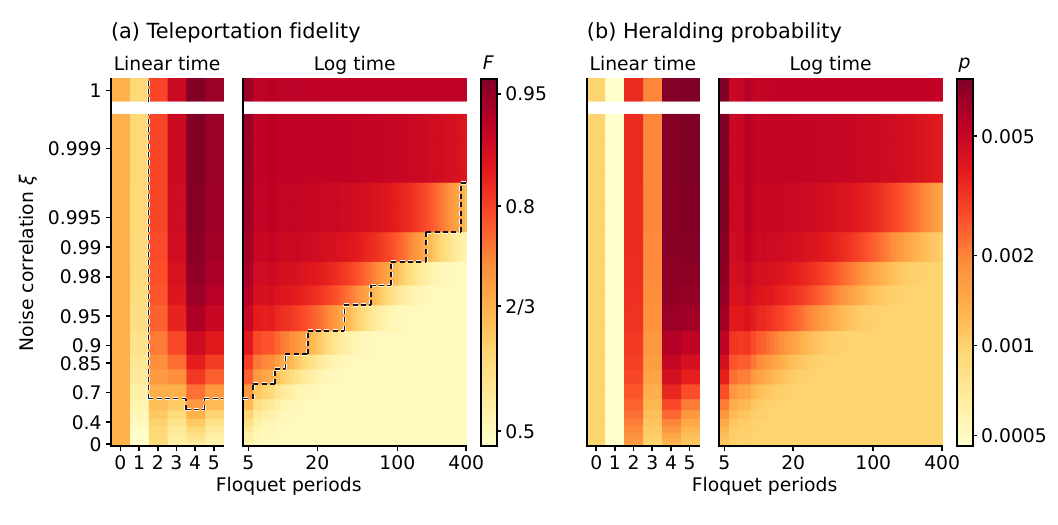}
\caption{\label{fig:dynamics}Conditional teleportation fidelity (a) and heralding probability (b) for the twelve-spin ring. The dashed boundary in (a) marks the sampled region with fidelity above the classical benchmark $2/3$. The probability colorbar in (b) uses a logarithmic scale. Each cell represents one integer Floquet period and one sampled correlation. In each panel, the narrow plot at left shows periods $0$ to $5$ on a linear time axis. The plot at right shows periods $5$ to $400$ on a logarithmic time axis. The correlation coordinate is $-\log_{10}(1-\xi)$, with the collective endpoint $\xi=1$ in a separate top row.}
\end{figure*}

The Lindblad generator $\mathcal D_\xi$ describes the dephasing layer, while $\mathcal M_\xi$ in Eq.~\eqref{eq:map} describes the full period. For collective dephasing, $\xi=1$, the stationary state here satisfies $\mathcal D_1[\rho_\infty]=0$ and $\mathcal M_1[\rho_\infty]=\rho_\infty$. Let $D_L$ comprise the ring's $L$ rotations and $L$ reflections, represented by spin permutations $U(g)$. Its group average gives
\begin{equation}
\Psym=\frac{1}{2L}\sum_{g\in D_L}U(g),
\quad N_L=\Tr\Psym,\quad
\rho_\infty=\frac{\Psym}{N_L}.
\label{eq:target}
\end{equation}
Here $\Psym$ projects onto $\mathcal S$, the spin states unchanged by every rotation and reflection of the ring, and $N_L=\dim\mathcal S$. Thus $\rho_\infty$ is maximally mixed on this sector, which contains the uniform product input. Both coherent layers and collective dephasing preserve $\mathcal S$ and leave $\rho_\infty$ unchanged. Appendix~\ref{App:A} proves that every initial state supported on $\mathcal S$ converges to this density for twelve spins. Evaluating Eq.~\eqref{eq:negativity} on $\rho_\infty$ gives $\EN(\infty)\simeq0.942$, the plateau in Fig.~\ref{fig:negativity}. For $\xi<1$, additional noise modes resolve relative spin configurations, breaking this support protection and making $\rho_\infty$ nonstationary.

The next step is to extract a pair of usable spins. Concentrating many-body entanglement onto selected spins is the goal of localizable-entanglement protocols~\cite{Popp2005}. Our circuit combines gates and measurements within each half to prepare a pair for teleportation~\cite{Teleportation1999}. In each half of $m=L/2$ spins, select the alternating string $a=0101\cdots$ and its complement $\bar a$. The operator for a successful local outcome is
\begin{equation}
K=\ket0\bra a+\ket1\bra{\bar a}.
\label{eq:decoder}
\end{equation}
Alice and Bob each have access to one half of the spin chain, with six spins in each half for the twelve-spin system considered here. Within each half, apply five controlled-NOT gates, using the first spin as the control and each of the other five spins as a target. Measure the five target spins in the computational basis, leaving the first spin unmeasured. The selected strings, for six spins, $010101$ and $101010$ both produce the measurement record $10101$, leaving the retained spin in $0$ and $1$, respectively. Since the measurement does not distinguish these two configurations, their coherence is preserved in the retained spin. Alice and Bob communicate their measurement outcomes and keep the retained pair only if each obtains the predetermined record $10101$. All gates act within the individual halves, so the protocol extracts entanglement already present between them.

For a ring state $\rho_t$, let $\sigma_t$ be the unnormalized state selected by the joint measurement record. Its trace $p_t$ is the probability of obtaining that record. Conditioned on that record, the pair has state $r_t$.
\begin{equation}
\sigma_t=(K\otimes K)\rho_t(K^\dagger\otimes K^\dagger),\quad
p_t=\Tr\sigma_t,\quad r_t=\sigma_t/p_t.
\label{eq:pair}
\end{equation}
For the stationary state at $\xi=1$, Appendix~\ref{App:B} derives the accepted pair and its heralding probability.
\begin{equation}
r_\infty=\frac67\ket{\Phi^+}\!\bra{\Phi^+}+\frac17\ket{\Psi^+}\!\bra{\Psi^+},\qquad p_\infty=\frac1{192},
\label{eq:pair12}
\end{equation}
Here $\ket{\Phi^+}=(\ket{00}+\ket{11})/\sqrt2$ and $\ket{\Psi^+}=(\ket{01}+\ket{10})/\sqrt2$. The overlap with the reference Bell state is $f\equiv \bra{\Phi^+}r_\infty\ket{\Phi^+}=6/7$. Using this pair in the standard teleportation protocol, the output fidelity averaged over uniformly distributed pure input qubits is~\cite{Teleportation1999}
\begin{equation}
F=\frac{2f+1}{3}=\frac{19}{21}\simeq90.5\%>\frac23.
\label{eq:fidelity}
\end{equation}
The classical benchmark $2/3$ is the best mean fidelity when Alice measures the input and Bob prepares an output from her classical message, with no shared entanglement~\cite{Teleportation1999}. Equation~\eqref{eq:fidelity} gives the fidelity conditioned on a successful extraction. Alice and Bob herald success before receiving the unknown input. With independent resets, the steady-state yield $p_\infty=1/192$ means they need $192$ preparation attempts on average. We assume ideal local gates and readout, with the driven noisy dynamics paused during extraction and teleportation.

{\it Discussion and Conclusions.}---Figure~\ref{fig:dynamics}~(a) shows how noise correlations control the survival of the accepted pairs' teleportation advantage. At period zero, before Floquet evolution, a successful extraction from the initial product state retains the pair $\ket{++}$. Its squared overlap $f=1/2$ with $\ket{\Phi^+}$ gives $F(0)=(2f+1)/3=2/3$ for every $\xi$, setting the baseline at the classical benchmark. For sufficiently strong correlations, the mean fidelity rises above the benchmark after entanglement builds up. As correlations approach the collective limit, the fidelity stays above this benchmark for longer. For the strongest correlations, it survives to the end of the $400$-period run. Figure~\ref{fig:dynamics}~(b) gives the corresponding extraction probability. Ring entanglement can outlast the teleportation advantage of this decoder, distinguishing entanglement survival from its usefulness for the chosen task. Our simple, unoptimized filter serves as a proof of principle: despite its small success probability, it converts noise-protected many-body entanglement into a pair for high-fidelity teleportation. We focus on twelve spins to keep full density-matrix evolution and entanglement calculations tractable.

The trend follows what the environment can resolve. For $\xi<1$, additional noise channels distinguish differences between the spins. Their strength is proportional to $1-\xi$ (Appendix~\ref{app:Log-neg}), so this source of dephasing weakens as $\xi\rightarrow1$. At the collective endpoint, the environment couples only to the sum of all spins' $z$ components, the total magnetization, and ring symmetry protects the entangled stationary state.

We show how interactions and collective noise prepare entanglement that can be used for teleportation. The twelve-spin dynamics approaches an entangled stationary state within the ring-symmetric sector. A circuit acting within the two halves extracts a physical pair with uniform-input average fidelity $19/21$ and heralding probability $1/192$. Superconducting circuits already realize kicked-Ising dynamics~\cite{Mi2022edge} and engineered correlated dephasing~\cite{vonLupke2020}. Building on our shared-reservoir circuit proposal~\cite{Ayoub2026}, we are pursuing experimental tests of correlated-noise effects, with entanglement extraction as the next goal. The protocol connects a many-body state protected by collective noise to a concrete quantum-information task.

\begin{acknowledgments}
\emph{Acknowledgements.}---This work was supported by the National Science Foundation (CHE-2404788) and the Robert A. Welch Foundation (E-1337). GPT-6.1 Sol assisted with the manuscript editing. The authors take full responsibility for the content.
\end{acknowledgments}

\emph{Data availability.}---
The numerical calculations are publicly available in the \href{https://github.com/bhavaytyagi/extracting-operational-entanglement-from-a-noise-protected-many-body-system}{GitHub repository}~\cite{Tyagi2026Data}.

\bibliographystyle{apsrev4-2}
\bibliography{References_revision_2}

\clearpage

\section*{End Matter}
\appendix
\newcommand{\prlappendix}[1]{\par\addvspace{4pt}\refstepcounter{section}\noindent\textit{Appendix~\thesection: #1}---}

\prlappendix{Convergence under collective dephasing}
\label{App:A}
Every state supported on $\mathcal S$, the ring-symmetric sector defined after Eq.~\eqref{eq:target}, converges to $\rho_\infty$ under $\mathcal M_1$ at the twelve-spin parameters of Eq.~\eqref{eq:map}. The coherent layers and the collective spin operators $J^\alpha=\tfrac12\sum_j\sigma_j^\alpha$ preserve $\mathcal S$. We restrict all operators below to this sector.

\emph{Which operators survive dephasing?} Consider an eigenoperator $\mathcal M_1[X]=\lambda X$ with $|\lambda|=1$. Set $U_F=\UK\UI$ and $Y=U_F X U_F^\dagger$. Collective dephasing leaves equal-magnetization blocks of $Y$ unchanged and strictly damps all other blocks ($\gamma>0$, Appendix~\ref{app:Log-neg}). Yet $\|e^{\mathcal D_1}(Y)\|_2=\|X\|_2=\|Y\|_2$ in the Hilbert--Schmidt norm. Hence $Y=\lambda X$ and $[J^z,X]=0$. Conjugating this commutator by $U_F$ shows that $X$ also commutes with $\UK J^z\UK^\dagger=\cos(2b)J^z-\sin(2b)J^y$, since $[\UI,J^z]=0$. At $b=-\pi/4$ this gives commutation with $J^y$, and then with $J^x$ by the spin commutator. Thus $X$ commutes with the kick and obeys $\UI X\UI^\dagger=\lambda X$. Conversely these conditions suffice, giving
\begin{align}
\operatorname{Fix}(\mathcal M_1)&=\{J^x,J^y,J^z,\UI\}',\nonumber\\
\operatorname{Per}_\lambda(\mathcal M_1)&=\{X:[X,J^{x,y,z}]=0,\ \UI X\UI^\dagger=\lambda X\}.
\label{eq:algebra}
\end{align}
Here the prime denotes the common commutant.

\emph{Uniqueness on $\mathcal S$.} The polarized vector $v=\ket{0\cdots 0}$ spans the highest-$J^z$ eigenspace. Every fixed operator $C$ therefore satisfies $Cv=cv$. To show that $C$ is scalar throughout $\mathcal S$, it suffices to span that sector with vectors $Av$ formed from the generators in Eq.~\eqref{eq:algebra}. For each such $A$, $CAv=cAv$.

At $J=\pi/4$, the periodic Ising layer is $\UI=e^{-i\pi L/4}R e^{-i2hJ^z}$, where $R=(-1)^{n_{\mathrm{DW}}/2}$ and $n_{\mathrm{DW}}$ counts domain walls. Thus $R$ and the joint spectral projectors of $R$ and $J^z$ belong to the generated algebra. Take as basis vectors the unnormalized sums of computational strings in each dihedral orbit. In this basis the joint projectors are diagonal with entries zero or one, and $2J^x=\sum_j\sigma_j^x$ has integer entries that count single-spin flips. Starting with $v$, we close its span under the projectors and $2J^x$ by retaining each linearly independent vector they generate. The rank is $224=N_{12}$ modulo each of the primes $1000003$ and $1000033$. A full-rank minor nonzero modulo either prime is nonzero over $\mathbb C$. Hence the vectors $Av$ span $\mathcal S$, and the fixed operators are scalar there.

\emph{Convergence.} For any eigenoperator in Eq.~\eqref{eq:algebra}, $X^\dagger X$ commutes with all three collective spin components and with $\UI$, since $|\lambda|=1$. Hence $X^\dagger X$ is scalar, and a nonzero $X$ is invertible on $\mathcal S$. Commutation with $J^z$ gives $Xv=cv$. Evaluating $\UI X\UI^\dagger=\lambda X$ on the Ising eigenvector $v$ gives $c=\lambda c$. Invertibility forces $c\ne0$, so $\lambda=1$. The Hilbert--Schmidt contraction bounds all powers of $\mathcal M_1$, excluding nontrivial Jordan blocks on the unit circle. All other modes decay, and trace preservation fixes the limiting density to Eq.~\eqref{eq:target}.

\prlappendix{Heralded pair from the stationary state}
\label{App:B}
The decoder in Eq.~\eqref{eq:decoder} selects four computational strings. Their orbits determine the accepted pair without constructing the full density matrix. If $\mathcal O_y$ is the dihedral orbit of $y$, each string in that orbit occurs equally often in the group average of Eq.~\eqref{eq:target}. Thus
\begin{equation}
\bra x\Psym\ket y=\begin{cases}1/|\mathcal O_y|,&x\in\mathcal O_y,\\0,&x\notin\mathcal O_y.\end{cases}
\label{eq:orbits}
\end{equation}
For $a=010101$, $aa$ and $\bar a\bar a$ are the two translates of the alternating string. The string $a\bar a=010101101010$ has a single $11$ bond, so its twelve translates are distinct. Reversing the string leaves it unchanged and translating by six sites gives $\bar a a$. The decoder maps the selected strings from the first orbit to $00,11$ and those from the second to $01,10$. Within each selected block all entries of $\Psym$ are equal, $1/2$ in the first and $1/12$ in the second. As $N_{12}=224$, Eq.~\eqref{eq:pair} gives
\begin{equation}
\sigma_\infty=\frac{1}{224}\left(\ket{\Phi^+}\!\bra{\Phi^+}
+\frac{1}{6}\ket{\Psi^+}\!\bra{\Psi^+}\right).
\end{equation}
Its trace gives $p_\infty=(1+1/6)/224=1/192$. Dividing by this probability gives the Bell weights $6/7$ and $1/7$ in Eq.~\eqref{eq:pair12}. With $\Phi^+$ as the reference and all teleportation outcomes retained, the first weight gives the conditional mean fidelity $19/21$ through Eq.~\eqref{eq:fidelity}.

\prlappendix{Noise modes and correlation range}
\label{app:Log-neg}
For the generator in Eq.~\eqref{eq:noise}, write the covariance of the Gaussian phase increments as $\mathbb E[dW_i\,dW_j]=\gamma\Xi_{ij}\,dt$, with
\begin{equation}
\bm\Xi=(1-\xi)\mathbb I+\xi\bm e\bm e^{\mathsf T},
\qquad \bm e=(1,\ldots,1)^{\mathsf T}.
\label{eq:noise_covariance}
\end{equation}
The normalized uniform mode $\bm v_0=\bm e/\sqrt L$ has eigenvalue $\lambda_0=1+(L-1)\xi$. The $L-1$ orthogonal modes have zero component sum and eigenvalue $\lambda_\perp=1-\xi$. Nonnegativity of both mode strengths gives
\begin{equation}
\bm\Xi\succeq0\quad\Longleftrightarrow\quad
-\frac{1}{L-1}\le\xi\le1.
\label{eq:noise_range}
\end{equation}
Relative to independent noise, $\xi<0$ weakens the uniform mode and strengthens the differential modes. At $\xi=-1/11$ for twelve spins, the phase increments sum to zero.

Write $\mathfrak D[A]\rho=A\rho A^\dagger-\tfrac12\{A^\dagger A,\rho\}$. For a real orthonormal eigenbasis $\bm v_\alpha$, averaging the phase rotations generated by $\sum_j\sigma_j^z dW_j$ gives Eq.~\eqref{eq:noise} in terms of the jump operators $L_\alpha$,
\begin{align}
\mathcal D_\xi[\rho]&=\sum_{\alpha=0}^{L-1}\mathfrak{D}[L_\alpha]\rho,
\nonumber\\
L_\alpha&=\sqrt{\gamma\lambda_\alpha}
\sum_j v_\alpha(j)\sigma_j^z.
\label{eq:noise_jumps}
\end{align}
Each differential mode therefore enters the dissipator with strength $\gamma(1-\xi)$. At $\xi=1$ only the uniform mode remains. Its eigenvalue $L$ cancels the squared normalization of $\bm v_0$, giving
\begin{equation}
L_0=\sqrt\gamma\sum_j\sigma_j^z=2\sqrt\gamma J^z,
\qquad \mathcal D_1[\rho]=\mathfrak D[L_0]\rho.
\label{eq:collective_jump}
\end{equation}
{\widowpenalty=10000 Coherences within a magnetization eigenspace then survive the dephasing step, as used in Appendix~\ref{App:A}. For example, the two alternating strings in Appendix~\ref{App:B} have the same total magnetization but opposite local spins. Their coherence in $\rho_\infty$ is damped whenever $\xi<1$, because their spin difference lies entirely in the differential subspace. The coherent layers preserve $\rho_\infty$, so this damping makes it nonstationary.\par}

\newpage
\prlappendix{Finite-time observables}
The trajectories in Figs.~\ref{fig:negativity} and~\ref{fig:dynamics} use the full twelve-spin density at every correlation, including $\xi=1$. At all $401$ integer times we check trace and Hermiticity to $10^{-10}$ and obtain the extraction probability and conditional fidelity from the four strings in Appendix~\ref{App:B}.

For the logarithmic negativity in Eq.~\eqref{eq:negativity}, we take the partial transpose on the first six spins and evaluate its full spectrum at periods $0,1,2,3,4,5,10,20,40,80,150,250,400$. At these times we also check density positivity and both spectral traces to $10^{-10}$, using the computed eigenvalues without clipping. GPT-6-Astra assisted with all numerical calculations and suggested the extraction-protocol.
\end{document}